\documentclass[10pt,journal,compsoc,fleqn]{IEEEtran}
\newcommand{\qed}{\hfill\IEEEQED}
\usepackage[nocompress]{cite}
\usepackage{xspace}
\usepackage{url}
\usepackage{xcolor}
\usepackage{soul}
\usepackage{amsfonts}
\usepackage{amsmath}
\usepackage{amssymb}
\usepackage{z-eves}
\usepackage{booktabs}
\usepackage{tabularx}
\usepackage{dcolumn}
\usepackage{colortbl}
\usepackage{tikz}
\usetikzlibrary{positioning,fit,calc,babel,backgrounds,
                shapes,arrows,automata,decorations.pathmorphing}

\newtheorem{example}{Example}

\newcommand{\setlog}{$\{log\}$\xspace}

\renewcommand{\Cup}{\mathsf{un}}
\renewcommand{\Cap}{\mathsf{inters}}
\newcommand{\In}{\mathbin{\mathsf{in}}}
\newcommand{\Dom}{\mathsf{dom}}
\newcommand{\Subseteq}{\mathsf{subset}}
\newcommand{\Comp}{\mathsf{comp}}
\newcommand{\Ncomp}{\mathsf{ncomp}}
\newcommand{\Apply}{\mathsf{apply}}
\newcommand{\Pfun}{\mathsf{pfun}}
\newcommand{\Or}{\mathbin{\mathsf{or}}}
\newcommand{\Ncup}{\mathsf{nun}}
\newcommand{\Nin}{\mathbin{\mathsf{nin}}}
\newcommand{\Ris}{\mathsf{ris}}
\newcommand{\Neq}{\mathbin{\mathsf{neq}}}
\newcommand{\Forall}{\mathsf{foreach}}

\def\Optional{\mathop{\rm optional}}
\def\The{the~}%
\def\Nil{nil}%

\begin{document}
\title{An Automatically Verified Prototype of the Tokeneer ID Station Specification}

\author{Maximiliano~Cristi\'a~and~Gianfranco~Rossi%
\IEEEcompsocitemizethanks{\IEEEcompsocthanksitem M. Cristi\'a is  with Universidad Nacional de Rosario and CIFASIS, Argentina.
\protect\\
E-mail: \texttt{cristia@cifasis-conicet.gov.ar}
\IEEEcompsocthanksitem G. Rossi is with Universit\'a di Parma, Italy.
\protect\\
E-mail: \texttt{gianfranco.rossi@unipr.it}}%
\thanks{to be completed}}

\markboth{to be completed}%
{Cristi\'a and Rossi: An Automatically Verified Prototype of the Tokeneer ID Station Specification}

\IEEEtitleabstractindextext{%
\begin{abstract}
The Tokeneer project was an initiative set forth by the National Security
Agency (NSA, USA) to be used as a demonstration that developing highly secure
systems can be made by applying rigorous methods in a cost effective manner.
Altran Praxis (UK) was selected by NSA to carry out the development of the
Tokeneer ID Station. The company wrote a Z specification later implemented in
the SPARK Ada programming language, which was verified using the SPARK Examiner
toolset. In this paper, we show that the Z specification can be easily and
naturally encoded in the \setlog set constraint language, thus generating a
functional prototype. Furthermore, we show that \setlog's automated proving
capabilities can discharge all the proof obligations concerning state
invariants as well as important security properties. As a consequence, the
prototype can be regarded as correct with respect to the verified properties.
This provides empirical evidence
that Z users can use $\setlog$  to generate correct prototypes from their Z
specifications. In turn, these prototypes enable or simplify some verification
activities discussed in the paper.

\end{abstract}

}

\maketitle

\IEEEdisplaynontitleabstractindextext

\section{Introduction}\label{sec:introduction}
\IEEEPARstart{F}{ormal} methods (FM) are still questioned with respect to their
actual value to deliver software at a reasonable cost. Formal methods
researchers and practitioners have proved many times that formal methods can
deliver software of unmatched quality, e.g.
\cite{DBLP:journals/cacm/Leroy09,DBLP:conf/sp/MurrayMBGBSLGK13}. The FM
community has also shown that high quality is not an impediment to keep costs
low when total cost of ownership and critical systems are considered
\cite{DBLP:journals/tse/KingHCP00,DBLP:journals/software/HallC02}. However,
most of the software industry is still reluctant to apply FM, and frequently is
unaware of their possible value. Even critical software providers do not always apply FM and are not obligated
to do so as standards do not always mandate a formal approach, e.g. IEC 61508
Safety Integrity Level 4.

In this context the National Security Agency (NSA) of the
U.S.A. conducted a project aiming at providing evidence to software vendors
that formal techniques can deliver high quality software in time and within
budget. In particular NSA chose the Tokeneer ID Station (TIS) which makes part
of the Tokeneer system. Tokeneer provides protection to secure information held
on a network of workstations situated in a physically secure enclave. TIS, in
turn, is a stand-alone trusted entity responsible for performing biometric
verification of the user. NSA asked Altran Praxis (then Praxis Critical
Systems) to provide an implementation of TIS conforming to Common Criteria's
EAL5-level \cite{cc315} and to disclose to the public domain all the
deliverables produced during system development. As said, the ultimate goal of
NSA was to show that this kind of development efforts are feasible (i.e., they
achieve reliable software) and cost-effective (i.e., they are not more
expensive than traditional development processes)
\cite{Barnes00,tokennerSumRep}.

Altran Praxis applied its own Correctness by Construction development process
to the TIS software. The second phase of this development process consists in
writing a Z \cite{Spivey00,Potter} formal specification of the user requirements. The
Z specification is central to the development process as it is used as the
correctness criteria for many verification activities.

The work described in this paper starts from this Z specification. More
precisely, we first \emph{encode} the Z specification in the \setlog set
constraint programming language
\cite{Dovier00,DBLP:journals/jar/CristiaR20,setlog}. This provides a functional
prototype of the Z specification. We say `encode' and not `implement' due to
the close resemblance between the \setlog language and Z; however, the encoding
provides an implementation in the form of a prototype. In a second step, we use
\setlog's constraint solving capabilities to automatically prove that the
prototype verifies \emph{all the state invariants} defined in the Z
specification as well as \emph{all but one of the security properties} stated
by Altran Praxis team---this amounts to discharge 523 proof obligations. This
constitutes a step forward with respect to the original project as these
invariants and properties were not machine-checked by Altran Praxis at the
specification level. Besides, this implies that the \setlog program is an
automatically verified functional prototype of the Z specification.
This provides empirical evidence that Z users can use $\setlog$  to generate
correct prototypes from their Z specifications. The paper discusses some
verification activities that can be carried out or simplified once the \setlog
prototype is available.

The present work aims at providing more empirical evidence that FM-based tools such as \setlog can
effectively be used in industrial projects willing to provide high quality
software without incurring in increased costs or delayed schedules.

The paper assumes the reader has some exposure to Z specification---otherwise
the reader can consult any Z textbook, e.g. \cite{Potter}.

The paper is structured as follows. The Z specification of the Tokeneer project
is briefly introduced in Section \ref{tisspec}, along with the security
properties the system should enforce. \setlog is presented in Section
\ref{setlog} by means of examples. Section \ref{encoding} discusses the
encoding of the Z specification in \setlog. Section \ref{prototype} shows how
the \setlog prototype is verified by running automated proofs of state
invariants and security properties; Section \ref{veriftool} discusses some
other verification activities that can be done with the prototype. A
quantitative report of some key aspects of this endeavor is presented in
Section \ref{quantrep}. In Section \ref{relwork} we put our results in the
context of other works that used the Tokeneer project as a case study. In
Section \ref{concl} we present our conclusions.

\section{\label{tisspec}The Z Specification of the TIS}
The Tokeneer project carried out be Altran Praxis has been thoroughly
documented in part due to the requirements of the Common Criteria
\cite{tokeneer,Barnes00}. In this section we will focus on the Z specification
generated during the project. The goal is for the reader to have an idea of the
complexity and peculiarities of the Z specification. We will not explain it in
full in part because, precisely, encoding it in \setlog can be done in a
completely formal manner; understanding what the specification is about is unnecessary.

The Z specification is a 117 pages long document containing formal Z code plus
informal explanatory statements about it \cite{tokeneer-z}. There is also an 11
pages long document stating security properties the Z specification must verify
\cite{tokeneer-secprop}. Whenever we refer to ``the (Z) specification'' we mean
these two documents, unless stated differently; and by ``the team'' we mean the
Altran Praxis development team that worked out the specification.

The Z specification is written in a more or less standard fashion. However, it
starts by introducing two polymorphic operators working in tandem that are
somewhat unusual:
\begin{zed}
\Optional X == \{x:\finset X | \# x \leq 1\} \\
the[X] == \{x:X @ \{x\} \mapsto x\}
\end{zed}
$\Optional$ is used to indicate that a variable can hold a value or nil, as for
example in:
\begin{schema}{Certificate}
        id: CertificateId
\\  validityPeriod: \power TIME
\\      isValidatedBy: \Optional KEYPART
\end{schema}
meaning that the certificate can be validated by some (asymmetric
cryptographic) key or it cannot. When the key does exist and
has to be retrieved from the certificate, then the $the$ operator comes
into play; for instance:
\begin{zed}
\The cert.isValidatedBy = issuerCert.subjectPubK
\end{zed}
Concerning the encoding of the Z specification in \setlog, it is important to
note that $\Optional$ can be defined as follows:
\begin{zed}
\Optional X == \{x:\finset X | x = \emptyset \lor (\exists y:X | x = \{y\})\}
\end{zed}
Likely, the team opted by the first definition in an attempt to keep the number
of quantified formulas as low as possible. As we will show, introducing
existentially quantified variables in \setlog is harmless, while introducing
the cardinality operator is not.

Z schemas are heavily used to give structure to the main concepts formalized in
the specification. For example, $Certificate$ is used to define:
\begin{schema}{AttCertificate}
        Certificate
\\      baseCertId: CertificateId
\\      tokenID: TOKENID
\end{schema}
which in turn is used to define:
\begin{schema}{PrivCert}
    AttCertificate
\\  role: PRIVILEGE
\\  clearance: Clearance
\end{schema}
which is used to define a $Token$:
\begin{schema}{Token}
    tokenID: TOKENID

\also   idCert: IDCert
\\  privCert: PrivCert
\\  iandACert: IandACert
\\  authCert: \Optional AuthCert
\end{schema}

It can be said that the specification is divided into two parts: the real world
peripherals interacting with the TIS and the TIS itself. The state of all real
world entities is modeled with the $RealWorld$ schema which is divided into two
schemas, $TISControlledRealWorld$ and $TISMonitoredRealWorld$. In total
$RealWorld$ comprises 11 state variables. In turn, the TIS state is modeled in
schema $IDStation$ which includes 12 schemas representing different subsystems.
In total $IDStation$ declares 36 state variables. The definition of $IDStation$
follows the style of most Z specifications\footnote{We use ellipses to shorten
the presentation.}:
\begin{schema}{IDStation}
    \dots
\\      DoorLatchAlarm
\\  \dots
\\  currentDisplay: DISPLAYMESSAGE
\\      currentScreen: Screen
\where
    status \in \{gotFinger, waitingFinger,\dots~\} \implies
\\ \t1       (( \exists ValidToken @
            goodT(\theta ValidToken) = \dots)
\\ \t2  \lor ( \exists TokenWithValidAuth @
            \dots))
\\
\dots
\\   
        currentScreen.screenStats = displayStats (\theta
        Stats)
\end{schema}
That is, it includes several schemas, declares two variables and 8 state
invariants. Note that these state invariants are conjoined with those declared
in some of the included schemas (e.g. $DoorLatchAlarm$). This implies a total
of 13 state invariants. The reader can see two
of them in the summary of the schema shown above. The first one gives an idea
of the complexity of some of the predicates: it is a quantified formula using
Z's $\theta$ operator. The $\theta$ operator is one of the most complex logical
operators of the Z notation.

Some of the state variables in $RealWorld$ and $IDStation$ have a
non-enumerated free type, as for example:
\begin{zed}
        KEYBOARD ::= noKB | badKB | keyedOps \ldata \dots \rdata
\end{zed}
which means the presence of structured infinite types.

The specification defines 25 major operations plus 3 that group some of these
operations (e.g. $TISUserEntryOp$ specifying the complete authentication
process)\footnote{These are the operation schemas whose names begin with
$TIS$.}. All TIS operations are state transitions over $RealWorld$ and
$IDStation$. Many of these operations are assembled from simpler operation
schemas by means of some non-trivial schema expressions; for instance:
\begin{zed}
TISArchiveLogOp \defs StartArchiveLog \lor \dots \also
StartArchiveLog \defs \\
\t1 (StartArchiveLogOK \semi UpdateFloppy)
\\ \t1                  \lor StartArchiveLogWaitingFloppy
\\ \t1  \lor
        [~ BadAdminLogout | \dots ~]
\end{zed}

There is a particularly important operation schema, namely $TISProcessing$,
specifying the overall processing activity of the TIS because it is used to
state some important security properties as we will see. $TISProcessing$
``calls'' 20 TIS operations and updates the audit log:
\begin{zed}
TISProcessing \defs \\
  \t1 (\dots \lor TISUserEntryOp \lor \dots
   \lor TISAdminOp \lor \dots) \\
  \t1 \land LogChange
\end{zed}

\subsection{\label{tisstateinv}State Invariants}
The specification follows the Z style concerning the encoding of state
invariants. This means that state invariants are declared in the predicate part
of the state schema (i.e., $IDStation$). Including the state invariants in the
state schema implies that all operations trivially verify them. We exemplify
this with a simple example not taken from the specification.
\begin{zed}
State \defs [x:\num | 0 \leq x] \also
Op \defs [\Delta State | x' = x - 1]
\end{zed}
$Op$ cannot violate the state invariant ($0 \leq x$) because the state
invariant is part of $Op$' definition:
\begin{zed}
[\Delta State | x' = x - 1] \\
\equiv [x,x':\num | 0 \leq x \land x' = x - 1 \land 0 \leq x']
\end{zed}
So $Op$
cannot be called when $x = 0$ because $Op$'s predicate implies $0 < x$. This is
called \emph{implicit precondition} and raises due to the interaction between
the state invariant and the operation predicate.

Another approach concerning state invariants is to prove that each operation
verifies them. In this case the state invariant is not encoded in the state
schema, preconditions are explicitly stated and a proof obligation is
introduced:
\begin{zed}
State \defs [x:\num] \also
Inv \defs [State | 0 \leq x] \also
Op \defs [\Delta State | 0 < x \land x' = x - 1] \also
\textbf{proof obligation } Inv \land Op \implies Inv'
\end{zed}

This second approach has the advantage that all preconditions are
explicit making the transition to the implementation simpler, but it has
the disadvantage of having to discharge proof obligations.

In encoding the Z specification in \setlog we opted for this second approach
because we want the \setlog prototype to be closer to an implementation (than
the Z specification) and because \setlog can automatically discharge these
proof obligations (see Section \ref{prototype}).

\subsection{\label{secprop}Security Properties}
The team stated 6 security properties the specification must verify; one of
them is not formalized, so we will not consider it. The team provided informal
proofs of these properties. According to the documentation the most important
of these properties is Property 1; we reproduce an excerpt of it here for the
reader to have an idea of its complexity:
\begin{zed}
\Delta IDStation; \Delta RealWorld |
\\ \t1  TISOpThenUpdate
\\ \t1  \land latch = locked \land latch' = unlocked
\\ \shows
\\ \t1  (\exists ValidToken @
            goodT(\theta ValidToken) = \dots
\\ \t2      \land UserTokenOKNoCurrCheck
\\ \t2      \land FingerOK
        )
\\ \t1  \lor
\\ \t1  (\exists TokenWithValidAuth @
            goodT(\dots) = \dots
\\ \t2      \land UserTokenWithOKAuthCertNoCurrCheck
        )
\\ \t1  \lor
\\ \t1      (\exists ValidToken @
            goodT(\theta ValidToken) = \dots
\\ \t2          \land authCert \neq \emptyset
                \land (\The authCert).role = guard
        )
\end{zed}
where:
\begin{zed}
    TISOpThenUpdate \defs TISProcessing \semi TISUpdate
\end{zed}
That is, the property must be proved for 20 of the TIS operations, followed by
the $TISUpdate$ operation.

It is important to remark that the above formula does not really correspond to
the property the team would have wanted to prove---we call it the
\emph{intended property}. This is acknowledged by team in the technical
documentation \cite[page 11, `Note on arguments']{tokeneer-secprop}. The
formula formalizing the intended property cannot be expressed in the Z
notation. Indeed, that formula would have to predicate over (infinite)
sequences of states instead of pairs of states as stated by the semantics of
the Z notation \cite{DBLP:conf/zum/Evans97,DBLP:conf/zum/Lamport94}. It is easy
to observe that TIS is actually a reactive system perhaps making Z not the best
notation to specify it.

We remark this point because we make an extra effort to overcome this limitation
when proving the validity of Property 1 for the \setlog prototype (see Section
\ref{prop1}). Hence, we end up producing not only a mechanized (automated)
proof of Property 1 but also we produce proofs of several other properties that
together provide much stronger arguments that the intended property
actually holds.

\section{\label{setlog}The \setlog Constraint Solver}
\setlog is a public available satisfiability solver and a set-based,
constraint-based programming language implemented in Prolog \cite{setlog}.
\setlog implements a decision procedure for the theory of \emph{hereditarily
finite sets}, i.e., finitely nested sets that are finite at each level of
nesting \cite{Dovier00}; a decision procedure for a very expressive fragment of
the class of finite set relation algebras
\cite{DBLP:journals/jar/CristiaR20,DBLP:conf/RelMiCS/CristiaR18}; a decision
procedure for restricted intensional sets (RIS)
\cite{DBLP:conf/cade/CristiaR17,DBLP:journals/corr/abs-1910-09118}; and uses
Prolog's CLP(Q) to provide a decision procedure for integer linear arithmetic
\cite{DBLP:conf/cp/HolzbaurMB96}. In \setlog sets and binary relations are
first-class entities of the language. At the core of these decision procedures
is set unification \cite{Dovier2006}. The set terms defined in all these three
decision procedures can be combined in several ways: binary relations are
hereditarily finite sets whose elements are ordered pairs and so set operators
can take binary relations as arguments; RIS can be passed as arguments to set
operators and freely combined with extensional sets. \setlog is an untyped
formalism; variables are not declared; typing information can be encoded by
means of constraints. Several in-depth empirical evaluations provide evidence
that \setlog is able to solve non-trivial problems
\cite{DBLP:journals/jar/CristiaR20,DBLP:conf/RelMiCS/CristiaR18,DBLP:conf/cade/CristiaR17,DBLP:journals/corr/abs-1910-09118,CristiaRossiSEFM13},
in particular as an automated verifier of security properties
\cite{Cristi__2020}. Given that \setlog has been extensively described
elsewhere, in this section we will show a few examples for the reader to
understand how it works.

In \setlog set operators are encoded as constraints. For example: $\Cup(A,B,C)$
is a constraint interpreted as $C = A \cup B$. \setlog implements a wide range
of set and relational operators covering most of those used in Z. For instance,
$\In$ is a constraint interpreted as set membership (i.e., $\in$); $=$ is set
equality; $\Dom(F,D)$ corresponds to $\dom F = D$; $\Subseteq(A,B)$ corresponds
to $A \subseteq B$; $\Comp(R,S,T)$ is interpreted as $T = R \comp S$ (i.e.,
relational composition); and $\Apply(F,X,Y)$ is equivalent to $\Pfun(F) \And
[X,Y] \In F$, where $\Pfun(F)$ constrains $F$ to be a (partial) function.
Formulas in \setlog are conjunctions (\&) and disjunctions ($\Or$) of
constraints; they must finish with a dot (as a Prolog query). Negation in
\setlog is introduced by means of so-called \emph{negated constraints}. For
example $\Ncup(A,B,C)$ is interpreted as $C \neq A \cup B$ and $\Nin$
corresponds to $\notin$---in general, a constraint beginning with
`$\mathsf{n}$' identifies a negated constraint. For formulas to lay inside the
decision procedures implemented in \setlog, users must only use this form of
negation.

The fact that set operators take a
relational rather than a functional form makes it necessary to
introduce variables to write compound expressions.
\begin{example}\label{ex:expressions}
The Z expression
$x \in A \cap B \cap C$
is encoded as
$X \In M \And \Cap(A,B,W) \And \Cap(W,C,M)$.
\qed
\end{example}
Set terms can be of the following forms:
\begin{itemize}
  \item A variable is a set term; variable names must start with an uppercase letter.
  \item $\{\}$ is the term interpreted as the empty set.
  \item $\{x / A\}$ is called \emph{extensional set} and is interpreted as
$\{x\} \cup A$; $A$ must be a set term, $x$ can be any term accepted by \setlog
(basically, any Prolog uninterpreted symbol, integers, lists, ordered pairs,
etc.).
  \item $\Ris(X \In A, \phi)$ is called \emph{restricted intensional set}
(RIS) and is interpreted as $\{x : x \in A \land \phi\}$ where $\phi$ is any
\setlog formula; $A$ must be a set term and $X$ is a bound variable local to
the RIS. Actually, RIS have a more complex and expressive structure
\cite{DBLP:conf/cade/CristiaR17,DBLP:journals/corr/abs-1910-09118}.
\end{itemize}

Being a satisfiability solver, \setlog can be used as an automated theorem
prover. To prove that formula $\phi$ is a theorem, \setlog has to be called to
prove that $\lnot\phi$ is unsatisfiable.

\begin{example}
We can prove that set union is commutative by asking \setlog to prove the
following is unsatisfiable:
\begin{gather*}
\Cup(A,B,C) \And \Cup(B,A,D) \And C \Neq D.
\end{gather*}
As there are no sets satisfying this formula \setlog answers \textsf{no}. Note
that the formula can also be written with the $\Ncup$ constraint: $\Cup(A,B,C)
\And \Ncup(B,A,C)$. \qed
\end{example}

\setlog is also a programming language at the intersection of declarative
programming,  set programming \cite{DBLP:books/daglib/0067831} and constraint
programming.
 \setlog programs can be structured by means
of \emph{clauses}---as in Prolog. A clause can be seen as a subroutine or
procedure. Clauses can receive zero or more arguments. The only way a clause
can return a value is by means of one or more of its arguments. Under certain
conditions clauses behave as formulas. That is a \setlog clause can be seen as
both a program and a formula. The following examples show the
\emph{formula-program duality} of \setlog code along with the notion of clause.

\begin{example}\label{ex:update}
If we want a program that updates function $F$ in $X$ with value $Y$ provided
$X$ belongs to the domain of $F$ and get an error otherwise, the \setlog code
can be the following:
\begin{gather*}
\mathsf{update}(F,X,Y,F\_,Error) \text{ :-} \\
\quad F = \{[X,V]/F1\} \And [X,V] \Nin F1 \And \\
\quad F\_ = \{[X,Y]/F1\} \And \\
\quad Error = ok \\
\quad \Or \\
\quad \Comp(\{[X,X]\},F,\{\}) \And \\
\quad Error = err.
\end{gather*}
That is, \textsf{update} is a clause that receives $F$, $X$ and $Y$ and returns
the modified $F$ in $F\_$ and the error code in $Error$---think of $F\_$ as the
value of $F$ in the next state. As $\&$ and $\Or$ are logical connectives and
$=$ is logical equality, the order of the `instructions' is irrelevant w.r.t.
the functional result---although it can have an impact on the performance.
Variable $F1$ is an existentially quantified variable representing the `rest'
of $F$ w.r.t. to $[X,V]$. If $[X,V]$ does not belong to $F$ then the
unification between $F$ and $\{[X,V]/F1\}$ will fail thus making
$\mathsf{update}$ to execute the other branch.

Now we can call \textsf{update} by providing inputs and waiting for outputs:
\[
\textsf{update}(\{[setlog,5],[hello,earth],[tokeneer,model]\},\\
\t2 hello,world,G,E).
\]
returns:
\begin{gather*}
G = \{[hello,world],[setlog,5],[tokeneer,model]\} \\
E = ok \tag*{$\qed$}
\end{gather*}
\end{example}

Since \textsf{update} is also a formula we can prove properties true of it.

\begin{example}
If $Error$ is equal to $err$ then $X$ does not belong to the domain of $F$. In
order to prove this property we need to call \setlog on its negation:
\[
\textsf{update}(F,X,Y,F\_,err) \And \Dom(F,D) \And X \In D.
\]
Then, \setlog answers \textsf{no} because the formula is unsatisfiable.
\qed
\end{example}

As Example \ref{ex:update} shows, variables introduced in the clause body (e.g. $F1$) are
existentially quantified variables. This is important because of the way
existentially quantified formulas of the Z specification are encoded in
\setlog.

\setlog implements set unification but syntactic Prolog
unification is still available as part of it. Prolog
unification comes handy to encode some Z features and predicates.
\begin{example}\label{ex:dotnotation}
The `dot' notation used in Z to access components of ordered pairs and
variables of schema types, can be encoded by means of Prolog unification. For
instance if $x:A \times B$, then the Z expression: $h = x.1$ can be encoded as
$X = [X1,X2] \And H = X1$ or even as $X = [H,X2]$. Further, since $X2$ seems to
be uninteresting yet another encoding is $X = [H,\_]$. \qed
\end{example}

\begin{example}\label{ex:indom}
A more elaborated example is the encoding of the Z predicate $x \in \dom f$ as
$[X,\_] \In F$. Indeed, $[X,\_] \In F$ is readily rewritten as $F = \{[X,\_] /
F1\}$ which is interpreted as `there is a pair in $F$ whose first component is
$X$', which in turn means that $X$ belongs to the domain of $F$. A more direct
encoding is $\Dom(F,D) \And X \In D$ but it requires to compute the domain
while the first encoding does not, meaning that the first encoding will, in
general, yield more efficient \setlog code. Yet another encoding is
$\Ncomp(\{[X,X]\},F,\{\})$ which uses the negated constraint of the $\Comp$
constraint. That is, $\Ncomp(\{[X,X]\},F,\{\})$ is equivalent to
$\Comp(\{[X,X]\},F,F1) \And F1 \Neq \{\}$, which is a fourth encoding. \qed
\end{example}

As can be seen, frequently, unification introduces existentially quantified
variables. In many circumstances these quantified formulas can be dealt with in
a decidable manner
\cite{Dovier00,DBLP:journals/jar/CristiaR20,DBLP:conf/RelMiCS/CristiaR18,DBLP:conf/cade/CristiaR17,DBLP:journals/corr/abs-1910-09118}.
A key aspect to preserve decidability is not to use logical negation but the
negated constraints provided by \setlog.
\begin{example}\label{ex:neg}
Concerning formula $x \in \dom f$ of Example
\ref{ex:indom}, the encoding of its negation, $x \not\in \dom f$, in
$\setlog$ cannot be done simply with a formula such as $\forall Y: [X,Y] \Nin
F$ because this formula is outside the set of admissible $\setlog$
formulas. Conversely, the decidable way to handle this negation is either:
$\Dom(F,D) \And X \Nin D$; or $\Comp(\{[X,X]\},F,\{\})$.
\qed
\end{example}

The last example opens the issue of universal quantifiers in \setlog. In
\setlog universally quantified formulas are provided by mean of RIS. In effect,
the introduction of RIS in \setlog allows for the definition of
\emph{restricted universal quantifiers} (RUQ). In general, if $A$ is a set,
then a RUQ is a formula of the following form:
\[
\forall x \in A: \phi
\]
It is easy to prove the following:
\begin{equation}\label{e:prop1}
(\forall x \in A: \phi) \iff A \subseteq \{x : x \in A \land \phi\}
\end{equation}
Given that $\{x : x \in A \land \phi\}$ is the interpretation of $\Ris(X \In A,\phi)$, the r.h.s. of \eqref{e:prop1} can be expressed as the \setlog formula:
\[
\Subseteq(A,\Ris(X \In A,\phi))
\]
In \setlog we have defined the $\Forall$ constraint to make RUQ easier to write:
\[
\Forall(X \In A,\phi) \text{ :- } \Subseteq(A,\Ris(X \In A,\phi)).
\]

We use these features to encode the Z specification in \setlog, to
automatically prove invariance lemmas and properties, and to provide a correct
prototype of the Z specification in the form of a \setlog program.

\section{\label{encoding}Encoding the TIS Specification in \setlog}
In the introduction, we say `encoding' and not `implementing' the TIS
specification due to the close resemblance between the Z and \setlog languages.
That is, our point is that writing \setlog code from the specification is
considerably more  natural, evident and semantically equivalent than writing,
say, SPARK code. Then, it looks like more as an encoding than as an
implementation. In particular, \setlog implements all the logical, set and
relational operators used in the TIS specification. Furthermore, these
operators are not mere imperative implementations but real executable
mathematical definitions. That is, they behave as logical or mathematical
objects, as we have shown in Section \ref{setlog}. In other words, the \setlog
prototype is a formula \emph{quite} as the TIS specification is. We say `quite'
because of two reasons:
\begin{itemize}
\item The \setlog code can be used to \emph{compute} results. Then, \setlog
programmers may pay attention to implementation issues, such as efficiency.
\item For different programming reasons (e.g. use of Prolog and set unification)
we introduce some modifications in the \setlog code w.r.t. the specification.
\end{itemize}
This takes the discussion to another issue: the \setlog code is a
\emph{prototype} and not a \emph{program}. Hence, users cannot expect the same
computing efficiency from the \setlog code than from a typical imperative
implementation. For instance, engineers can use the prototype to analyze
functional scenarios but they cannot draw efficiency estimations. This section will make these points clear.

The \setlog encoding of the TIS specification as well as all the proof
obligations can be found online: \url{http://people.dmi.unipr.it/gianfranco.rossi/SETLOG/APPLICATIONS/tokeneer.zip}.

In general, each Z schema is encoded as a \setlog clause. The variables
declared in the schema become arguments of the clause. In many cases, if the
schema declares variables through schema inclusion then the arguments of the
clause are those schemas instead of the variables declared inside them. This
somewhat preserves the structure of the specification. Schemas whose predicate
part is empty are encoded as tuples of variables. We preserved the identifiers
used in the specification as much as possible in the \setlog code. Recall that
\setlog variables (constants) must start with an uppercase (lowercase) letter;
while clauses can start only with a lowercase letter. Next state variables
(e.g. $x'$) are encoded as \setlog variables decorated with an underscore (e.g.
$X\_$)\footnote{The prime symbol is not allowed as part of a variable name in
Prolog+\setlog.}.

Figures \ref{f:encoding1} and \ref{f:encoding2} show parts of the encoding. In
Figure \ref{f:encoding1} the specification is at the left and the corresponding
encoding in \setlog, at the right. We attempted to align each row of the
specification with the corresponding row in \setlog. In Figure
\ref{f:encoding2} the specification is at the top and the encoding at the
bottom. In general, the encoding in \setlog is longer than the corresponding Z
code, as is expected for any lower-level representation, but not that much (see
Section \ref{quantrep}).

Consider Figure \ref{f:encoding1}. As can be seen, schema $CurrentToken$ is
encoded as clause \verb+currentToken+. $CurrentToken$ declares several
variables through the inclusion of schema $ValidToken$ which in turns declares
those variables through the inclusion of schema $Token$. Then, in this case
\verb+currentToken+ has two arguments: \verb+Token+, corresponding to schema
$Token$; and \verb+Now+, corresponding to variable $now$. In order to make
\verb+Token+ a valid one, \verb+validToken(Token)+ is conjoined. As
expected, \verb+validToken+ is the encoding of schema $ValidToken$ (not
shown). In the specification $TIME$ is a synonym for $\nat$, so \verb+Now+ is
constrained to be a non-negative integer by asserting \verb+0 =< Now+. Given
that $Token$ is a schema declaring five variables it is encoded as a 5-tuple:
 \verb+Token = [_,IDC,PC,IAC,_]+
where the correspondence between Z and \setlog variables is implemented by
declaration order. Thus, \verb+PC+ corresponds to the third variable declared
in $Token$, i.e., $privCert$. In \verb+currentToken+ only \verb+IDC+, \verb+PC+
and \verb+IAC+ play some role, so the other two are hidden by putting
underscores in their positions. The named variables can have any name because
they are existentially quantified inside the clause. Besides,
 \verb+Token = [_,IDC,PC,IAC,_]+ forces the unification of the actual argument
to be a 5-tuple; in case this is not true the clause will fail. The
specification variables corresponding to \verb+IDC+, \verb+PC+ and
 \verb+IAC+
are of schema types (e.g. $privCert:PrivCert$, see Section \ref{tisspec}), so
they are encoded as tuples, e.g. \verb+PC = [_,PVP,_,_,_,_,_]+. In this
particular case \verb+PVP+ corresponds to the Z expression
$privCert.validityPeriod$ (recall Example \ref{ex:dotnotation}). Therefore, the
predicate stated in $CurrentToken$ is encoded as shown in Example
\ref{ex:expressions}.

\begin{figure*}
\begin{tabular}{p{.5\textwidth}p{.5\textwidth}}
\begin{schema}{CurrentToken}
\also
\also
\also
\also
\also
\also
\also
\also
\also
    ValidToken \\
now: TIME
\where
\also
\also
    now \in idCert.validityPeriod
\\ \t1      {} \cap privCert.validityPeriod
\\ \t1      {} \cap iandACert.validityPeriod
\end{schema}
\begin{schema}{UserTokenOK}
        KeyStore
\\      UserToken
\\      currentTime : TIME
\where
\also
\also
\also
    currentUserToken \in \ran goodT
\also
\also
\\  \exists CurrentToken @
\also
\also
\also
\also
\also
\also
\also
\also
\also
\also
\\ \t1      (
        goodT(\theta ValidToken) = currentUserToken
\\ \t1      \land now = currentTime
\\ \t1          \land (\exists IDCert @ \theta IDCert = idCert \land CertOK )
\\ \t1          \land (\exists PrivCert @ \theta PrivCert = privCert
\land CertOK )
\\ \t1          \land (\exists IandACert @ \\
    \t3 \theta IandACert =
iandACert \land CertOK )
                )
\end{schema}
&
\begin{verbatim}
currentToken(Token,Now) :-
  Token = [_,IDC,PC,IAC,_] &
    IDC = [[_,VP,_],[_,_]] &
    PC = [_,PVP,_,_,_,_,_] &
    IAC = [_,IVP,_,_,_,_] &

  validToken(Token) &
  0 =< Now &

  Now in M2 &
    inters(VP,PVP,M1) &
    inters(M1,IVP,M2).


userTokenOK(
    KeyStore,
    UserToken,CurrentTime,Now) :-
  0 =< CurrentTime &
  UserToken = [CurrentUserToken,_] &

  CurrentUserToken = goodT(_) &

  currentToken(Token,Now) &
    Token = [_,IDC,PC,IAC,_] &
      IDC = [[Id,VP,IVB],[_,_]] &
      PC = [PCId,PCVP,PCIVB,_,_,_,_] &
      IAC = [IAId,IAVP,IAIVB,_,_,_] &

    goodT(Token) = CurrentUserToken &
    Now = CurrentTime &
    certOK(KeyStore,[Id,VP,IVB]) &
    certOK(KeyStore,[PCId,PCVP,PCIVB]) &
    certOK(KeyStore,[IAId,IAVP,IAIVB]).

\end{verbatim} \\
\end{tabular}
\caption{\label{f:encoding1}Comparison between Z code and \setlog code (part
1). Similarities are not only syntactic but semantic as well.}
\end{figure*}

Still in Figure \ref{f:encoding1}, $UserTokenOK$'s predicate is based on a
quantified schema expression using the $\theta$ operator. The
type of $currentUserToken$ is:
\begin{zed}
    TOKENTRY ::= noT | badT | goodT \ldata Token \rdata
\end{zed}
Then, $currentUserToken \in \ran goodT$ means that the current user token is
indeed a $Token$. That is, the predicate is equivalent to $\exists t:Token @
currentUserToken = goodT(t)$. Avoiding quantified formulas is important but the
encoding can introduce an existentially quantified variable without getting
into troubles: \verb+CurrentUserToken = goodT(_)+. Along the same lines, the
predicate $(\exists CurrentToken @ \dots)$ is easily encoded with
\verb+currentToken(Token,Now)+; that is, by asserting the existence of
variables \verb+Token+ and \verb+Now+ satisfying \verb+currentToken+. Then,
$\theta ValidToken$ refers to the $Token$ asserted in $(\exists CurrentToken @
\dots)$. So, \verb+goodT(Token) = CurrentUserToken+ is asserted in \setlog.
This implies that \verb+goodT(_) = goodT(Token)+, thus making
\verb+CurrentUserToken = goodT(_)+ superfluous---actually, we included it to
make the encoding more similar to the specification. Finally, the last three
existentially quantified predicates are encoded along the same lines. That is,
the encoding of schema $CertOK$ is called on each certificate contained in the
$Token$. Note that, for instance, $(\exists IDCert \dots)$ corresponds to
variable \verb+IDC+; thus $\theta IDCert = idCert$ corresponds to
 \verb+IDC = [[Id,VP,IVB],[_,_]]+; and \verb+[Id,VP,IVB]+ is the $Certificate$
passed in as argument to \verb+certOK+.

Now consider Figure \ref{f:encoding2}. Schema $BioCheckRequired$ specifies part
of the $ValidateUserTokenOK$ operation which in turn is part of the
$TISValidateUserToken$ operation. $BioCheckRequired$ makes a `delta' on
$IDStation$ and $RealWorld$; this is hidden in schema $UserEntryContext$. This
is reflected in the encoding as clause \verb+bioCheckRequired+ waits for
\verb+IDStation+ and \verb+IDStation_+, meaning that the clause
transitions from the former to the latter. As explained
in Section \ref{tisspec}, $IDStation$ includes 12 schemas and declares two
variables. Then, the encoding for $IDStation$ is a 14-tuple respecting the
declaration order given in the specification. As mentioned in Section
\ref{tisstateinv}, the invariants stated inside $IDStation$ are moved out of
it; see Section \ref{prototype} for further details. Note that $IDStation'$ is
encoded as \verb+IDStation_+.

As $BioCheckRequired$ states $\Xi UserToken$ the encoding forces
\verb+UserToken+ and \verb+UserToken_+ to be the first elements of
\verb+IDStation+ and \verb+IDStation_+, respectively, while
 \verb+UserToken_ = UserToken+ is conjoined to the clause.

$AddElementsToLog$ is an operation schema making a `delta` on the auditing
subsystem, $AudiLog$, and accessing the configuration subsystem, $Config$. In
\setlog this schema corresponds to clause \verb+addElementsToLog+. The clause
is called inside a \verb+delay+ predicate for efficiency reasons. After the
first proof obligations were discharged it was evident that
\verb+addElementsToLog+ was taking too long while not contributing to the
proofs. Hence, we ask \setlog to delay its execution as much as possible.
$AddElementsToLog$'s predicate starts by asserting the existence of a finite
set of audit records, $newElements$. The \setlog encoding reflects that
fact by declaring variable \verb+_NewElements+ inside the clause and
passing it to \verb+addElementsToLog+. This simplifies the
encoding of some tricky parts of the specification.

In $BioCheckRequired$, $status$ is a variable declared in schema $Internal$
(which is one of the 12 schemas included in $IDStation$). The encoding uses
unification to access the corresponding variable, \verb+Status+. Indeed,
unification forces \verb+Internal+ to be the 12th component of \verb+IDStation+
and it is used again to force \verb+Status+ to be the first component of
\verb+Internal+. Then, we can state \verb+Status = gotUserToken+. Note how the
change of state of $status$ is encoded by asserting
 \verb+Status_ = waitingFinger+.

\begin{figure*}
\begin{schema}{BioCheckRequired}
        UserEntryContext
\also
        \Xi UserToken
\\      \Xi DoorLatchAlarm
\\      \Xi Stats
\also
        AddElementsToLog
\where
        status = gotUserToken
\\      userTokenPresence = present
\also
        \lnot UserTokenWithOKAuthCert \land UserTokenOK
\also
    currentDisplay' = insertFinger
\\  status' = waitingFinger
\end{schema}

\begin{zed}
ValidateUserTokenOK \defs BioCheckRequired  \lor BioCheckNotRequired
\end{zed}

\begin{verbatim}
  bioCheckRequired(IDStation,RealWorld,RealWorld_,IDStation_) :-
    IDStation = [UserToken,_,_,DoorLatchAlarm,_,_,Config,Stats,
                 KeyStore,_,AuditLog,Internal,_,CurrentScreen] &
      UserToken = [_,UserTokenPresence] & DoorLatchAlarm = [CurrentTime,_,_,_,_,_] &
      Internal = [Status,_,_] & CurrentScreen = [ScreenStats,_,ScreenConfig] &
    RealWorld = [_,TISMonitoredRealWorld] &
      TISMonitoredRealWorld = [Now,_,_,_,_,_,_] &
    IDStation_ = [UserToken_,_,_,DoorLatchAlarm_,_,_,_,Stats_,
                _,_,AuditLog_,Internal_,CurrentDisplay_,CurrentScreen_] &
      Internal_ = [Status_,_,_] & CurrentScreen_ = [ScreenStats_,_,ScreenConfig_] &

    userEntryContext(IDStation,RealWorld,RealWorld_,IDStation_) &

    UserToken_ = UserToken &
    DoorLatchAlarm_ = DoorLatchAlarm &
    Stats_ = Stats &

    delay(addElementsToLog(Config,_NewElements,AuditLog,AuditLog_),false) &

    Status = gotUserToken &
    UserTokenPresence = present &

    delay(not_userTokenWithOKAuthCert(KeyStore,UserToken,CurrentTime),false) &
      userTokenOK(KeyStore,UserToken,CurrentTime,Now) &

    CurrentDisplay_ = insertFinger &
    Status_ = waitingFinger &

    ScreenStats_ = ScreenStats &
    ScreenConfig_ = ScreenConfig.

  validateUserTokenOK(IDStation,RealWorld,RealWorld_,IDStation_) :-
    bioCheckRequired(IDStation,RealWorld,RealWorld_,IDStation_)
    or
    bioCheckNotRequired(IDStation,RealWorld,RealWorld_,IDStation_).
\end{verbatim}

\caption{\label{f:encoding2}Comparison between Z code and \setlog code (part
2). Similarities are not only syntactic but semantic as well.}
\end{figure*}

The predicate $\lnot UserTokenWithOKAuthCert$ is encoded by calling the
negation of \texttt{userTokenWithOK} \texttt{AuthCert}. As explained in Section
\ref{setlog} (see Example \ref{ex:neg}), in order to preserve decidability
(general) logical negation should not be used. Instead, the negation has to be
written in terms of \setlog constraints. Given that Figure \ref{f:encoding1}
includes the encoding of $UserTokenOK$, Figure \ref{f:notusertokenok} shows the
encoding of $\lnot UserTokenOK$ (in place of $\lnot UserTokenWithOKAuthCert$,
which is nonetheless similar). There are several reasons for which the user
token is not OK. The first one occurs when the current user token does not
belong to the range of $goodT$. This implies that it must be either $noT$ or
$badT$ (see $TOKENTRY$ above). The encoding is
 \verb+CurrentUserToken in {noT, badT}+.
The other cases occur when there is a $token:Token$ such that $currentUserToken
= goodT(token)$. In this case, $(token,now)$ might not conform a current token
(\verb+not_currentToken(Token,Now)+); or $now$ might not coincide with
$currentTime$ (\verb+Now neq CurrentTime+); or any of the certificates stored
in $token$ is not OK (\verb+not_certOK(KeyStore,[_,_,_])+).

To close this section, note that the \setlog encoding of $ValidateUserTokenOK$
turns out to be quite 'natural' (both, syntactically and semantically). This
encoding becomes sort of a pattern by means of which most of the Z
specification is encoded. Considering the size and complexity of the TIS Z
specification, it can be argued that \setlog can be used as an effecitive
prototyping/programming language to encode many other Z specifications.

\begin{figure}
\begin{verbatim}
not_userTokenOK(KeyStore,UserToken,
                CurrentTime,Now) :-
  UserToken = [CurrentUserToken,_] &

  (CurrentUserToken in {noT, badT}
   or
   CurrentUserToken = goodT(Token) &
    Token = [_,IDC,PC,IAC,_] &
      IDC = [[Id,VP,IVB],[_,_]] &
      PC = [PCId,PCVP,PCIVB,_,_,_,_] &
      IAC = [IAId,IAVP,IAIVB,_,_,_] &
   (not_currentToken(Token,Now)
    or
    Now neq CurrentTime
    or
    not_certOK(KeyStore,[Id,VP,IVB])
    or
    not_certOK(KeyStore,[PCId,PCVP,PCIVB])
    or
    not_certOK(KeyStore,[IAId,IAVP,IAIVB])
   )
  ).
\end{verbatim}
\caption{\label{f:notusertokenok} Encoding of $\lnot UserTokenOK$}
\end{figure}

\subsection{Potential Problems with the Encoding}
In spite of the similarities between Z and \setlog, in passing from the
specification to the prototype several changes were introduced. We claim that, still, the prototype is a faithful representation of the specification as
it is possible to prove that the former verifies essential properties of the
latter---see Section \ref{prototype}. However, some of the changes might cause
troubles when proving new properties. In this section we give a brief account
of these differences.

\emph{Types are not always encoded.} Z is a typed formalism, \setlog is not.
Typing information can be encoded using set membership and constraints such as
$\Pfun$. We encoded the typing information needed to prove the properties
discussed in Section \ref{prototype}. Some typing information is given as state
invariants---as is nonetheless the case when, for instance, a set is
implemented as a list. The encoding assumes the Z specification has been
type-checked. Then, the encoding needs to check only the types of input values
and the initial state. It would be possible to provide a type-checker for
\setlog programs based on special-purpose typing constraints.

\emph{Elements beyond \setlog's expressiveness.} There are a few elements in
the specification that are beyond the expressiveness of \setlog. Schema types
cannot be encoded as sets of records. In this particular specification this
feature is used only once and can be circumvented. Free types declaring
non-constant elements (e.g. $TOKENTRY$ above) cannot be encoded as sets.
However, it is possible to assert that a variable is of that type (e.g.
\verb+userTokenOK+ and \verb+not_userTokenOK+), and that all the elements of a
set are of that type by means of the $\Forall$ constraint; not needed in
this specification. \setlog cannot express $\nat$ nor $\num$. It can
express that a variable belongs to them and that all the elements of a set
belong to them (this is used to encode the type of $sizeElement$). For this
reason, we cannot encode the type of $authPeriod$ and some values given in
$InitConfig$. It is doubtful how these values can be implemented in any
programming language as they entail to store a Cartesian product where one of
the sets is $\nat$.

\emph{Predicate outside the decision procedure.} The universally quantified
predicate in schema $ValidEnrol$ (not included for brevity) lays outside
the decision procedures implemented in \setlog---and evidence suggests that
this is a fundamental problem
\cite{DBLP:journals/jar/CristiaR20,andreka1997decision}. The problem is that
the property asserted for all elements in $issuerCerts$ depends on
$issuerCerts$ itself. In other words, for each element in $issuerCerts$ there
must exist another element in it fulfilling a certain property. This creates a
sort of recursive definition. However, the predicate \emph{can} be encoded and
it \emph{can} be used for running the prototype. Problems could arise if
\setlog is asked to decide the satisfiability of a formula involving it, when
$issuerCerts$ remains variable. In that case \setlog might enter an infinite
loop. This is apparently not the case for the set of properties considered in
the Tokeneer project.

\emph{Computationally hard predicate.} The predicate $oldElements \cup
auditLog' = auditLog \cup newElements$ in schema $AddElementsToLog$ is
computationally very hard when all the operands remain variable. We were able
to circumvent this issue by enclosing its encoding inside a \verb+delay+
predicate.

\medskip

As the empirical data shows (Section \ref{prototype}), these issues do not
threaten the chances of using \setlog as a verification tool for
industrial-strength Z specifications.

\section{\label{prototype}A Verified Prototype}
Now that we have the $\setlog$ program, we can use \setlog to prove properties
of it. We \emph{automatically}
prove two kinds of properties: invariance lemmas and security
properties. In any case, recall that \setlog can prove a theorem by proving
that its negation is unsatisfiable. If the theorem is of the form $p \implies
q$, then one should ask \setlog to check if $p \land \lnot q$ is unsatisfiable.
In doing so consider that: \emph{a)} the negation in $\lnot q$ should be
encoded as indicated in Sections \ref{setlog} and \ref{encoding}; and \emph{b)}
if $p \land \lnot q$ happens to be satisfiable, \setlog will produce a finite
representation of all the possible solutions (countermodels or
counterexamples). This last feature comes handy to find out what are the
possible causes for the formula not to be a theorem.

Section \ref{quantrep} provides quantitative figures about the verification
work carried out with \setlog.

\subsection{\label{lemmainv}Invariance Lemmas}
As explained in Section \ref{tisstateinv}, we move the state
invariants included in schema $IDStation$ out of it; this includes all the
state invariants written in the schemas included in $IDStation$. In doing so,
we must: \emph{a)} add pre- or post-conditions to some operations; and
\emph{b)} prove that each TIS operation preserves each invariant.

Concerning \emph{a)}, for instance, the encoding of $UnlockDoor$ is augmented
with the following conditions:

\begin{verbatim}
unlockDoor(DoorLatchAlarm,Config,
           DoorLatchAlarm_) :-
......
  (LatchUnlockDuration neq 0 &
   CurrentLatch_ = unlocked &
   DoorAlarm_ = silent
   or
   LatchUnlockDuration = 0 &
   CurrentLatch_ = locked &
   (AlarmSilentDuration neq 0 &
    DoorAlarm_ = silent
    or
    AlarmSilentDuration = 0 &
    (CurrentDoor = open &
     DoorAlarm_ = alarming
     or
     CurrentDoor = closed &
     DoorAlarm_ = silent
    )
   )
  ).
\end{verbatim}
for $TISUnlockDoor$ to preserve the invariant stated in schema $DoorLatchAlarm$:
\begin{zed}
    currentLatch = locked \iff currentTime \geq latchTimeout
\\  doorAlarm = alarming \iff
\\ \t1      (currentDoor = open
\\ \t2          \land currentLatch = locked
\\ \t2          \land currentTime \geq alarmTimeout
            )
\end{zed}
Note that the encoding of the invariant is closer to an implementation based on
conditional statements than the Z invariant and that it takes into account
other predicates included in $UnlockDoor$ like $latchTimeout' = currentTime +
latchUnlockDuration$. This would help in passing from the \setlog prototype to
a definitive implementation.

Concerning \emph{b)}, for each TIS operation we discharge a proof obligation of the form:
\begin{zed}
Invariant \land TISOperation \implies Invariant'
\end{zed}
encoded as:
\begin{verbatim}
invariant(IDStation) &
tisOperation(IDStation,RealWorld,
             RealWorld_, IDStation_) &
not_invariant(IDStation_).
\end{verbatim}
where \verb+not_invariant+ is the \setlog negation of \verb+invariant+.

\begin{example}
The state invariant included in $IDStation$:
\begin{zed}
enclaveStatus \\
  \t1 \notin \{~notEnrolled, waitingEnrol, waitingEndEnrol ~\} \\
    \t2 \implies  ownName \neq \Nil
\end{zed}
is encoded as:
\begin{verbatim}
idStationInv07(IDStation) :-
  IDStation =
    [...,KeyStore,...,Internal,...] &
    KeyStore = [_,OwnName] &
    Internal = [_,EnclaveStatus,_] &
  (EnclaveStatus in {notEnrolled,
                     waitingEnrol,
                     waitingEndEnrol}
   or
   OwnName neq {}
  ).
\end{verbatim}
while its negation is:
\begin{verbatim}
not_idStationInv07(IDStation) :-
  IDStation =
    [...,KeyStore,...,Internal,...] &
    KeyStore = [_,OwnName] &
    Internal = [_,EnclaveStatus,_] &
  EnclaveStatus nin {notEnrolled,
                     waitingEnrol,
                     waitingEndEnrol} &
  OwnName = {}.
\end{verbatim}
\qed
\end{example}

\subsection{\label{lemmasecprop}Security Properties}
All security properties formalized by the team---i.e., Property 1-4 and 6---are
encoded in \setlog. However, Property 2 cannot be proved by \setlog because it
requires a decision procedure for integer intervals---which is, in fact, part
of our current work. Properties 3, 4 and 6 are proved as formalized by the
team.
For instance, the encoding
of Property 3 is the following\footnote{The ellipsis replace the unification
between \texttt{IDStation} and the other arguments with tuples to have access
to the variables.}:
\begin{verbatim}
property3 :-
  idStationInv01(IDStation) &
  (tisEarlyUpdate(IDStation,RealWorld,
                  RealWorld_,IDStation_)
   or
   tisUpdate(IDStation,RealWorld,
             RealWorld_,IDStation_)
  ) &
......
  Latch_ = locked &
  CurrentDoor_ = open &
  CurrentTime_ >= AlarmTimeout &
  Alarm_ neq alarming.
\end{verbatim}

\subsubsection{\label{prop1}Proving Property 1}
As we have explained in Section \ref{secprop}, the formalization of Property 1
given by the team does not actually capture the intended property. This
is acknowledged by the team in the technical documentation. The intended
property requires to consider different execution sequences of the TIS
operations. Each of these sequences takes the system from the initial state to
a state where the property holds. The point made by the team is that the
system can follow only those sequences.

Therefore, we make one more step by proving that the system can only execute
those state sequences. Once the system arrives at the desired state, we prove
that Property 1 holds. Hence, the proof of Property 1 involves (automatically)
discharging 16 proof obligations---11 to prove the system can only execute
certain state changes; 4 to prove that some properties hold at some of the
traversed states; and 1 to prove Property 1. This proof strategy follows the
informal proof made by the team \cite[Section 3.2.1, pages
8-10]{tokeneer-secprop}.

Before giving details on the encoding of these proofs, it should be noted that
in doing them we have to add $enclaveStatus \neq waitingStartAdminOp$ as an
hypothesis to many of the auxiliary lemmas---otherwise they do not hold. This
is so because, without that hypothesis, $TISShutdownOp$ can be executed
violating some desired properties. To the best of our knowledge, this extra
hypothesis and the problems with $TISShutdownOp$ are not mentioned in the
documentation. Maybe this is obvious for the team and so they omitted it in the
documentation.

Figures \ref{f:statechanges1} and \ref{f:statechanges2} depict the state
sequences that lead the system to the state where Property 1 holds. Figure
\ref{f:property1_01} shows the encoding of a lemma stating that if the system
ever reaches $status = gotUserToken$ it is because the before state was $status
= quiescent$---i.e., the first transition of Figure \ref{f:statechanges1}. That
is, the \setlog code corresponds to the negation of the following formula:
\begin{zed}
\Delta IDStation; \Delta RealWorld | \\
  \t1 TISOpThenUpdate \\
  \t1 \land enclaveStatus \neq waitingStartAdminOp \\
  \t1 \land status' = gotUserToken \land status' \neq status \\
\vdash \\
  \t1  status = quiescent
\end{zed}
where $TISOpThenUpdate$ is the composition between the disjunction of all the TIS operations and the $TISUpdate$ operation \cite[page 5]{tokeneer-secprop}.

\begin{figure*}
\begin{tikzpicture}
[tran/.style={auto,->,shorten <=1pt,>=stealth',
             semithick,bend left=45},
 node distance=2.5cm,
 background rectangle/.style={fill=gray!7},
 show background rectangle]
\node[circle,minimum size=4pt,fill,inner sep=0pt] (ini) at (-1.25,-2) {};
\node[inner sep=2pt] (s0) at (0,-2) {$quiescent$};
\node[inner sep=2pt] (s1) at (2,0) {$gotUserToken$};
\node[inner sep=2pt] (s2) at (5,0) {$waitingFinger$};
\node[inner sep=2pt] (s3) at (7.75,0) {$gotFinger$};
\node[inner sep=2pt] (s4) at (10.75,0) {$waitingUpdateToken$};
\node[inner sep=2pt] (s5) at (14,0) {$waitingEntry$};
\node[draw,inner sep=2pt] (s6) at (14,-2) {$waitingRemoveTokenSuccess$};
\node[inner sep=2pt] (s7) at (2,-2) {$waitingEntry$};
\node[inner sep=2pt] (s8) at (0,-4) {$waitingRemoveTokenFail$};

\draw[->] (ini.east) -- (s0.west);
\draw[->] (s0.north) |- (s1.west);
\draw[->] (s1.east) to node {} (s2.west);
\draw[->] (s2.east) to node {} (s3.west);
\draw[->] (s3.east) to node {} (s4.west);
\draw[->] (s4.east) to node {} (s5.west);
\draw[->] (s5.south) to node {} (s6.north);
\draw[->] (s1.south) to node {} (s7.north);
\draw[->] (s7.east) to node {} (s6.west);
\draw[->] (s8.north) to node {} (s0.south);
\draw[->] (s6.south) |- (s8.east);
\draw[-] (s6.south) -- ++(0,-1) -| (s0.south);
\end{tikzpicture}
\caption{\label{f:statechanges1}Sequences of state changes leading to $status = waitingRemoveTokenSuccess$}
\end{figure*}
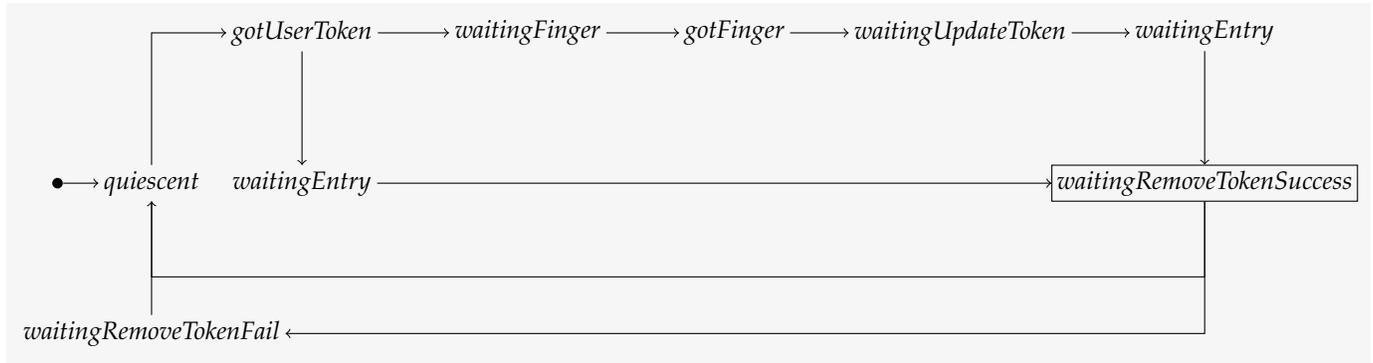

\begin{figure*}
\begin{tikzpicture}
[tran/.style={auto,->,shorten <=1pt,>=stealth',
             semithick,bend left=45},
 node distance=2.5cm,
 background rectangle/.style={fill=gray!7},
 show background rectangle]
\node[circle,minimum size=4pt,fill,inner sep=0pt] (ini) at (0,0) {};
\node[inner sep=2pt] (s0) at (2.5,0) {$enclaveQuiescent, \emptyset$};
\node[inner sep=2pt] (s1) at (6.5,0) {$gotAdminToken, \emptyset$};
\node[inner sep=2pt] (s2) at (11,0) {$enclaveQuiescent, \neq \emptyset$};
\node[draw,inner sep=2pt] (s3) at (15.75,0) {$waitingStartAdminOp, \neq\emptyset$};

\draw[->] (ini.east) -- (s0.west);
\draw[->] (s0.east) to node {} (s1.west);
\draw[->] (s1.east) to node {} (s2.west);
\draw[->] (s2.east) to node {} (s3.west);
\end{tikzpicture}
\caption{\label{f:statechanges2}Sequences of state changes leading to $enclaveStatus = waitingStartAdminOp, rolePresent \neq \emptyset$}
\end{figure*}
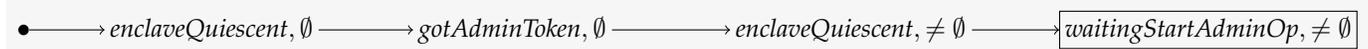

\begin{figure*}
\begin{verbatim}
  property1_01 :-
    IDStation = [_,_,_,DoorLatchAlarm,_,_,_,_,_,_,_,Internal,_,_] &
      Internal = [Status,EnclaveStatus,_] &
    EnclaveStatus = enclaveQuiescent &
    Status neq quiescent &
    tisOpThenUpdate(M,IDStation,RealWorld,RealWorld_,IDStation_) &
    IDStation_ = [_,_,_,_,_,_,_,_,_,_,_,Internal_,_,_] &
      Internal_ = [Status_,_,_] &
    Status neq Status_ & Status_ = gotUserToken.
\end{verbatim}
\caption{\label{f:property1_01}Encoding of a state change lemma}
\end{figure*}

We prove one such lemma for each transition shown in Figures
\ref{f:statechanges1} and \ref{f:statechanges2}. In this way, the \setlog
prototype is guaranteed to follow only those state sequences when it comes to
the validity of Property 1.

Next, we prove that when the system reaches some of the states depicted in
Figures \ref{f:statechanges1} and \ref{f:statechanges2} some properties hold.
For example, when passing from $gotUserToken$ to $waitingEntry$, the user token
is checked for validity and the presence in the token of an authorization
certificate is checked as well. Figure \ref{f:property1_09} shows the encoding
of this lemma.  As can be seen, we need \verb+pfun(IssuerKey_)+ as an
hypothesis. In Z terms, this means that $issuerKey'$ is a partial
function, which is the type given to the variable in the specification. This
hypothesis is proved to be a state invariant and so it can be assumed without
loss of generality.

\begin{figure*}
\begin{verbatim}
property1_09 :-
  IDStation = [_,_,_,_,_,_,_,_,_,_,_,Internal,_,_] &
    Internal = [Status,EnclaveStatus,_] &
  Status = gotUserToken & EnclaveStatus = enclaveQuiescent &
  tisOpThenUpdate(M,IDStation,RealWorld,RealWorld_,IDStation_) &
  IDStation_ = [UserToken_,_,_,_,_,_,_,_,KeyStore_,_,_,Internal_,_,_] &
    UserToken_ = [CurrentUserToken_,_] &
    Internal_ = [Status_,_,_] &
  Status_ = waitingEntry &
  (not_tokenWithValidAuth(CurrentUserToken_)
   or
   KeyStore_ = [IssuerKey_,_] &
   pfun(IssuerKey_) &
   not_userTokenWithOKAuthCertNoCurrencyCheck(KeyStore_,UserToken_,_)
  ).
\end{verbatim}
\caption{\label{f:property1_09}Encoding of a lemma stating an intermediate property}
\end{figure*}

Finally, Property 1 itself is proved by assuming the properties shown to be
valid in the intermediate states.

Having discharged all the proof obligations concerning state invariants and
those concerning security properties, the \setlog prototype is correct w.r.t.
them. Therefore, it can be used as a correct implementation of the
specification, at least on what concerns to the proven properties.

\subsection{\label{veriftool}The Prototype as a Verification Tool}
The prototype can be used as an analysis tool taking advantage of its
correctness. The most natural activity is to use it as an implementation to
evaluate how the system behaves in specific functional scenarios.
\begin{example}
We may want to analyze different scenarios when the enclave's door is unlocked
by asking \setlog to execute the following:
\begin{verbatim}
DLA = [5,locked,_,_,_,_] &
C = [10,4,_,_,_,_,_,_] &
doorLatchAlarmInv(DLA) & configInv(C,100) &
unlockDoor(DLA,C,DLA_).
\end{verbatim}
where part of the first solution is:
\begin{verbatim}
DLA_ = [5,locked,unlocked,silent,9,19]
\end{verbatim}
Observe that we give the before state (e.g. \verb+DLA+), check whether it satisfies the invariants, and then run \verb+unlockDoor+ waiting for the next state (\verb+DLA_+). Besides, note that the before state is only partially given (cf. the underscores in \verb+DLA+) which allows to analyze several scenarios in one run.
\qed
\end{example}
Being able to execute these scenarios in the presence of the customer can help
to ``check that this vital step [Z specification] has been achieved correctly''
without needing a person ``with full knowledge of the problem domain and a
fluent reading knowledge of Z'' \cite{Barnes00}--- the latter being not easy to
find.

A more elaborated activity would be to use the \setlog prototype as part of a
model-based testing method, e.g.
\cite{Stocks2,CristiaSTVR,DBLP:books/daglib/p/WoodcockAC10}. Let's call
$P_{\{log\}}$ and $P_{SPARK}$ the \setlog prototype and the SPARK
implementation corresponding to the Z specification, respectively. Given
a test case generated from the specification, run it on both $P_{\{log\}}$ and
$P_{SPARK}$ and compare the outputs generated by each of them to decide whether
or not the test case has uncovered an error in $P_{SPARK}$. The output
generated by $P_{\{log\}}$ should be deemed correct as the prototype verifies
all the stated properties. Hence, if the output generated by $P_{SPARK}$ does
not coincide it should be concluded that there is an error in it. Furthermore,
\setlog itself can be used as a test case generator in the context of Z
specifications \cite{CristiaRossiSEFM13}.

Along the same line, $P_{\{log\}}$ can be used as a runtime or reference
monitor of $P_{SPARK}$. Reference monitors have been proposed as a way to
control the execution of secure systems, e.g.
\cite{DBLP:journals/cuza/BetarteCLR16,DBLP:journals/cleiej/LunaBCSCG18}. In
this scenario, $P_{\{log\}}$ and $P_{SPARK}$ execute in parallel in such a way
that every input sent to $P_{SPARK}$ is also sent to $P_{\{log\}}$. If the
outputs differ an alarm is fired. Clearly, the control is \emph{ex post} (due
to the, possibly sensible, different execution speeds) but in some
circumstances this is much better than nothing. For example, in case a
malfunction in $P_{SPARK}$ allows an unauthorized access to the enclave, the
late alarm fired by $P_{\{log\}}$ would allow for security personnel to correct
the situation before is too late.

The intention behind these proposals and the mere fact of using \setlog is to
show that \setlog can be useful, and not that it should replace existing tools
or techniques used in development processes such as Altran Praxis' Correct by
Construction. \setlog can fill some gaps when it comes to prototyping,
automated proof and counterexample generation in the context of set-based
specifications.

\section{\label{quantrep}Quantitative Report}
Besides proving all the lemmas
described in Sections \ref{lemmainv} and \ref{lemmasecprop} we prove that:
\begin{enumerate}
\item Every disjunct of all the TIS operations is satisfiable.

This verification is important because, as operations act in the antecedent of
lemmas, if they are unsatisfiable then the lemma holds trivially. See for
example Property 1 in Section \ref{secprop}.
\item The initial state satisfies all the state invariants.

This is a standard verification step when state invariants are involved.
\item The conjunction of the negation of a schema and the schema itself
is unsatisfiable, while those schemas are independently satisfiable.

This applies only to those schemas whose negation appears either in the
specification or as part of a lemma. These proofs are done to gain confidence
in that each negation has been correctly written.
\end{enumerate}

Table \ref{t:summary} summarizes key quantitative measures about the effort
involved during this work. The first measured magnitude is effort (man-hours).
It is difficult and can be misleading to estimate the effort needed to conduct
the work carried out with \setlog because it was done on an academic
environment rather than a corporate one. Nevertheless, we think that an
approximate number could help others. Our effort estimate for encoding the Z
specification and discharging all the 523 proof obligations is between 40 and
120 man-hour. It depends on the experience of the engineer with Z, \setlog,
proof strategies and automated proof. Note that the effort reported by the team
to write the Z specification and conduct informal proofs is 233.5 man-hour
\cite[Appendix B, page 71, item 3000]{tokennerSumRep}. A more elaborated tool
environment could further reduce our estimate.

\newcolumntype{d}[1]{D{.}{.}{#1}}
\begin{table*}
\renewcommand{\arraystretch}{1.3}
\caption{\label{t:summary} Summary of Quantitative Report}
\centering
\begin{tabularx}{.5\textwidth}{Xd{3.1}d{3.1}}
\toprule
\rowcolor{gray!7}
\multicolumn{3}{c}{\textsc{Effort}} \\
\multicolumn{3}{l}{From 40 to 120 man-hour (prototype + proofs)} \\
\midrule
\rowcolor{gray!7}
\multicolumn{3}{c}{\textsc{Size}} \\
\rowcolor{gray!7}
&
\multicolumn{1}{c}{\textsc{KLOC}} &
\multicolumn{1}{c}{\textsc{KByte}} \\
\textsc{Z} (\LaTeX) & 2 & 52 \\
\textsc{Prototype} (\setlog) & 2.6 & 108 \\ 
\textsc{Lemmas} (\setlog) & 5.1 & 276 \\

\midrule
\rowcolor{gray!7}
\multicolumn{3}{c}{\textsc{Proof Obligations (Lemmas)}} \\
\rowcolor{gray!7}
\multicolumn{1}{c}{\textsc{Collection}} &
\multicolumn{1}{c}{\textsc{Number}} &
\multicolumn{1}{c}{\textsc{Time (s)}} \\
\textsc{Negations} & 93 & 4 \\
\textsc{Invariance} & 325 & 533 \\
$\t1 TISEnrolOp$ & & 107 \\
$\t1 TISValidateUserToken$ & & 114 \\
$\t1 TISPoll$ & & 281 \\
\textsc{Satisfiability of initial state} & 12 & 0.1 \\
\textsc{Satisfiability of operations} & 74 & 0.9 \\
\textsc{Security properties} & 19 & 313 \\
\midrule
\rowcolor{gray!7}
\textsc{Totals} & 523 & 851 \\
\bottomrule
\end{tabularx}
\end{table*}

The next measured magnitude is size. As can be seen, the \setlog prototype is
around 30\% larger than the Z specification measured in lines of code (LOC),
but it is more than the double in terms of kilobytes. This is consistent with
the fact that any implementation is expected to be more verbose than the
specification. This verbosity can be observed in Figures \ref{f:encoding1} and
\ref{f:encoding2}. The \setlog code for proof obligations (lemmas) is
considerably larger than the prototype, but it means less than 10 LOC per lemma
and it contains pretty-printing predicates to simplify proof execution. Much of
the proof obligations code can be automatically generated---we did so with a
few simple \verb+bash+ scripts but in an industrial environment it can be done
much better.

The last aspect we measure concerns the number of proof obligations (523) and
the time \setlog spends in discharging them (851 seconds). As can be seen, most
of the proof obligations corresponds to invariance lemmas as well as the
computing time to discharge them (548 seconds). Most of the time is spent in
proving the invariance of three TIS operations ($TISEnrolOp$,
$TISValidateUserToken$ and $TISPoll$). In average, each proof obligation runs
in 1.6 seconds. There are at least two possible ways to reduce the computing
time: parallelization and what we call \emph{proof programming}. In effect,
every lemma can be discharged independently of the others and so each of them
can be run in a different thread. Prolog provides high-level parallelization
predicates (e.g. \verb+concurrent/4+) that can be easily used to considerably
reduce the computing time.

Proof programming refers to the application of some \setlog control predicates
(e.g. \verb+delay+) to impose some order in the execution of constraints. As we
have shown in Section \ref{prototype}, enclosing a constraint in a \verb+delay+
predicate can speed up the execution of the prototype. However, if the delayed
constraint is important to discharge a particular proof obligation the proof
will take longer compared to a goal where that constraint is not delayed.
Hence, the best strategy is to delay a constraint for a proof while not to
delay it for another. This requires some \emph{proof programming}. In general
proof programming has to be applied only to some constraints. It is an area
that deserves to be further explored.

The numbers reported in Table \ref{t:summary} provide evidence about the
usefulness of \setlog as a verification tool for Z specifications.

\subsection{\label{details}Platform where the verification was executed}

The verification of the \setlog prototype of the TIS specification was
performed on a Latitude E7470 (06DC) with a 4 core Intel(R)
Core\texttrademark{} i7-6600U CPU at 2.60GHz with 8 Gb of main memory, running
Linux Ubuntu 18.04.4 (LTS) 64-bit with kernel 4.15.0-106-generic. \setlog
4.9.6-21c over SWI-Prolog (multi-threaded, 64 bits, version 7.6.4) was used
during the experiments.

The execution time of each collection of proof obligations is given by \verb+T+
in the following \setlog formula:
\begin{verbatim}
prolog_call(get_time(Ti)) &
<collection proof obligations>
prolog_call(get_time(Te)) &
prolog_call(T is Te - Ti).
\end{verbatim}
where \verb+prolog_call+ is a \setlog facility to access the Prolog interpreter.

The \setlog code used in this work can be downloaded from
\url{http://people.dmi.unipr.it/gianfranco.rossi/SETLOG/APPLICATIONS/tokeneer.zip} along with instructions to set up the environment.

\section{\label{relwork}Other works about the Tokeneer Project}

Besides members of the
team in charge of the Tokeneer project, other researchers have used it as a
case study, a benchmark or just as an industrial-scale problem. We will comment
on the most relevant ones w.r.t. our work.

Rivera et al. \cite{DBLP:conf/icse/RiveraBC16} undertake the Tokeneer project
in Event-B. They use the Rodin toolset for discharging proof obligations and
the EventB2Java code generator to create a Java program from the Event-B model.
The TIS model consists of an abstract machine and 6 refinements. The full
development resulted in 334 proof obligations, of which more than 90\% are
discharged automatically using Rodin. These proof obligations should coincide
with a subset of those proved with \setlog. Rivera and his colleagues prove
Properties 1, 2 and 3 with Rodin, although it is not clear whether or not they
are automatically discharged. As can be seen, \setlog is able to automatically
discharge 100\% of the proof obligations. Besides, \setlog produces a prototype
while the Event-B approach needs to apply EventB2Java. Likely, the resulting
Java program will be more efficient than the \setlog prototype, but the former
is \emph{only} a program while the latter is a program \emph{and} a formula.
Although it is not exactly the same, there is a sort of Java version of \setlog
called JSetL \cite{DBLP:conf/zum/CristiaR18,DBLP:journals/fuin/RossiB15}.

Answer Set Programming, which is close to constraint programming, has been used
to generate counterexamples for false and unprovable verification conditions
(VC) of the Tokeneer project \cite{DBLP:conf/iclp/SchandaB12}. These VC
correspond to those generated during the verification of the SPARK
implementation. As we have shown, \setlog returns a finite representation of
all the solutions of any satisfiable formula. These solutions are
counterexamples when the intention is to discharge a VC---that is, the VC is
negated, submitted to \setlog and it returns a solution (counterexample)
meaning that the VC is not a theorem. Along the same lines, in a technical
report, Jackson and Passmore apply an SMT solving-based tool to prove SPARK VC
of the Tokeneer project \cite{jacksonPassmore}. The tool calls CVC 3, Yices, Z3
and Simplify. Roughly, the tool proves more than 90\% of the VC. Tokeneer has
also been used as a case study for the formal verification framework Echo
\cite{DBLP:conf/nfm/YinK10}. Echo uses PVS as a theorem prover and SPARK as
programming language. According to the paper ``in 90\% of the cases, the PVS
theorem prover could not prove the implication lemmas completely
automatically''.

Although, the VC are at the SPARK level and the proof obligations discharged
by \setlog and Rodin (cf. Rivera et al. above) are at the specification level,
there should be a clear relation between them as the SPARK program should
implement the specification. Hence, \setlog looks promising as a VC verifier.

Abdelhalim et al. \cite{DBLP:conf/icfem/AbdelhalimSST10} apply CSP to formalize
fUML activity diagrams and FDR as a model checker to the Tokeneer
specification. Specifically the authors found several deadlock scenarios in the
form of counterexamples generated by FDR.

The work authored by Moy and Wallenburg \cite{moy2010tokeneer} is interesting
because they find problems in Tokeneer, although it was formally verified. Moy
and Wallenburg's goal is to find out why these problems were not found when the
system was verified and to propose verification activities that could have
detected these problems. Specifically, the authors propose to complement formal
verification with static analysis and code reviews. \setlog might be considered
as part of the toolbox proposed by Moy and Wallenburg as it is at the
intersection of several programming and verification paradigms. For instance,
it can be used to perform proofs and to run functional scenarios.

Woodcock at al. \cite{DBLP:books/daglib/p/WoodcockAC10} apply an
assertion-guided model-based robustness testing method to the Tokeneer project.
Robustness testing checks that a system can handle unexpected user input or
software failures. They use a model of the system for code generation (the Z
original specification) and a separate model for test case generation (an Alloy
model); these models are independently produced from the requirements. The test
case specifications are fed into the Alloy Analyzer, and test cases are
automatically generated as counterexamples. This allowed the authors to detect
nine anomalous behaviors. \setlog can be used in place of Alloy and its
analyzer. In fact the Alloy Analyzer does not implement a decision procedure
for sets and binary relations, as the one provided by \setlog. Then, the Alloy
Analyzer might fail in finding a counterexample while \setlog might not.

\section{\label{concl}Conclusions}
We have encoded the Z specification of the Tokeneer project in \setlog. This
encoding can be used as a functional prototype. Then, we used \setlog to
automatically proved hundreds of proof obligations over the prototype itself.
That is, we took advantage of the formula-program duality of \setlog code to
produce a verified prototype w.r.t. the proven properties. In this way, \setlog
is used as a programming language and a verification engine using the same and
only representation of the system. The case study provides evidence that
\setlog can be helpful in analyzing set-based formal specifications like those
written in the Z formal notation.

The most interesting future work is to apply \setlog to discharge the
verification conditions generated during the verification of the SPARK
implementation.


%




\bibliographystyle{IEEEtran}
\bibliography{/home/mcristia/escritos/biblio.bib}

%

\begin{IEEEbiography}{Maximiliano Cristi\'a}
to be completed.
\end{IEEEbiography}

\begin{IEEEbiography}{Gianfranco Rossi}
to be completed.
\end{IEEEbiography}





\end{document}